\documentclass{article}

\newtheorem{theorem}{Theorem}
\newtheorem{corollary}{Corollary}

\def \C{C \! \! \! \! 1}
\def\1{{\bf 1}}

\begin{document}
\setcounter{footnote}{1}

\title{Characterizations of the Generators of Positive Semi\-groups
on $C^{\ast}$- and von Neumann Algebras\footnote{Part of this work is contained in the author's thesis
\cite{Rheinlaender.97}.}}
\author{J\"org Rheinl\"ander \\
Institut f\"ur Mathematische Stochastik\\
Georg-August-Universit\"at \\ Lotzestr.\ 13, D-37083 G\"ottingen, Germany.\\
\small{e-mail: joerg.rheinlaender@gmx.net}}
\maketitle

\begin{abstract}
\noindent Generators of positive $ C_0$-semigroups on $C^{*}$-algebras and $
C_0^{\ast}$-semigroups on von Neumann
algebras are examined. A characterization due to Bratteli and Robinson in the $ C_0 $-case
is proven in the $ C_0^*$-case. Under the additional assumptions of unitality and
contractivity of the 
semigroup another characterization of the generator is given.
This result is restated for the dual and predual semigroup. 
\end{abstract}



\section[Introduction]{Introduction}
In this note we examine infinitesimal generators of positive one
parameter semigroups and of  positive, unital and contractive
one-parameter semigroups $ T_t $ on $ C^{\ast}$- and von 
Neumann algebras. Unitality of $ T_t $ means $ T_t(\1)=\1 $, where $ \1 $ is
the unit of the algebra.\\

Uniformly continuous, positive semigroups were studied by Evans and Han\-che-Olsen 
\cite{Evans.HancheOlsen.79}. They characterized the generators $ L $ of such semigroups 
by the inequalities
\begin{eqnarray*}
L(a^2)+aL(\1)a &\geq & L(a)a+aL(a), \qquad \mbox{for all self-adjoint} \quad a\in {\cal A}, \\
L(\1)+u^*L(\1)u &\geq & L(u^*)u+u^*L(u), \qquad \mbox{for all unitary} \quad u \in {\cal A}.
\end{eqnarray*}
This result was generalized by Bratteli and Robinson \cite{Bratteli.Robinson.81} to positive 
$ C_0$-semi\-groups on $ C^{\ast}$-algebras. They derived the above inequalities with the generator $ L $
replaced by the resolvent $ (\lambda-L)^{-1} $. In section 2 this result is restated for
$ C_0^{\ast}$-semigroups on von Neumann algebras. The proof of this characterization is
mainly based on the proof 
of Bratteli and Robinson in the $ C_0$-case. The modification of the
proof holds for both
$ C_0$-semigroups on $ C^{\ast}$-algebras and
$ C_0^{\ast}$-semigroups acting on von Neumann algebras.\\

In section 3 it is additionally assumed that the semigroups under
consideration are contractive and unital. 
A characterization
of the generators themselves can be given, which is not only a characterization in terms of their
resolvents. This result for positive, unital semigroups of 
contractions admits a reformulation 
for the related dual and predual semigroups. This leads to a generalization of a result
due to Kossakowski \cite{Kossakowski.72}.\\

With the help of the results in section 3 it is proven in \cite{Leschke.Rheinlaender.Schuetz.00}
that the generators of uniformly continuous, unital and positive semigroups of contractions
on the $ C^{\ast}$-algebra $ {\cal M}(n,\C) $ of $ n\times n $ matrices
are essentially the difference of two generators of
completely positive semigroups.\\
Semigroups which satisfy this stronger notion of positivity are the
natural candidates for a description of a Markovian irreversible time evolution in quantum
statistical mechanics, in particular for so called reduced dynamics.  
These dynamics arise from a Hamiltonian dynamic through an averaging
procedure over a subsystem or some internal degrees of freedom, i.e.\ one is interested
in the family of maps $ N\circ\alpha_t$, where $ \alpha_t $ is a one-parameter
group of $ *$-automorphisms of the algebra $ {\cal A} $ and $ N:{\cal A}\to {\cal B} $
is the conditional expectation from $ {\cal A} $ into a $ C^*$-subalgebra $ {\cal B} $. 
If $ N $ and $ \alpha_t $ are completely positive, so is the composition. Of course, 
$ N\circ \alpha_t $ is in general not a one-parameter semigroup because of memory
effects. In order to eliminate these memory effects, one usually carries out a limiting
procedure, for example a weak or singular coupling limit \cite{Duemcke.Spohn.79}.
For a detailed discussion see \cite{Evans.82}.\\
When the completely positive, unital semigroup is uniformly continuous, the infinitesimal
generator can be written in the so-called Lindblad form \cite{Gorini.Kossakowski.etal.76}, 
\cite{Lindblad.76}, \cite{Christensen.Evans.79}, see also \cite{Bratteli.Robinson.96} 
p. 222-226.
In the case of non-uniformly continuous semigroups, a canonical form of the generator is
not yet known, but there are interesting results in this direction, see \cite{Holevo.95},
\cite{Chebotarev.Fagnola.93} and references therein.\\
The description of irreversible dynamics by completely positive semigroups is not
without criticism, see e.g.\ \cite{Bratteli.Robinson.96}, \cite{Majewski.84} 
and \cite{Pechukas.94}.\\ 
For a detailed discussion on positive and completely positive semigroups 
see \cite{Leschke.Rheinlaender.Schuetz.00}.
 
\section[Positive semigroups]{Positive semigroups}
In this section we discuss positive semigroups and their
generators. We state a characterization of the generators
of both positive $ C_0$-semigroups on a $C^{\ast}$-algebra and positive
$ C_0^*$-semigroups on a von Neumann algebra ${\cal A}$, in the $C_0$-case due to
Bratteli and Robinson \cite{Bratteli.Robinson.81}.\\

Let us recall some definitions. If $ X $ is a Banach space then a one parameter
family $ T_t $, $ t\geq 0 $, of bounded linear operators on $ X $ is defined to be
a strongly continuous or $ C_0$-semigroup if it satisfies
\begin{enumerate}
\item $ T_tT_s=T_{t+s}, \quad \forall t,s \geq 0 $,
\item $ T_0 = id_X $,
\item $ \lim_{t \downarrow 0} \|T_t(x)-x\|=0, \quad \forall x \in  X $.
\end{enumerate}
Similarly, if $ X $ is a Banach space with predual space $ X_{\ast} $ a one parameter
family $ T_t $, $ t \geq 0 $, is defined to be a $ C_0^{\ast}$-semigroup if it satisfies
1 and 2 from above and the weak-$\ast$-continuity conditions
\begin{enumerate}
\item[4.] $ t \mapsto \eta(T_t(x)) $ is continuous $ \forall x \in X $, $ \forall \eta \in X_{\ast} $,
\item[5.] $ x \mapsto \eta(T_t(x)) $ is continuous $ \forall \eta \in X_{\ast} $, $ \forall t \geq 0 $.
\end{enumerate}
If one replaces the predual space $ X_{\ast} $ in the above definition by the dual space
$ X^{\ast} $ one gets a weakly continuous semigroup. It can be proven that each weakly continuous
semigroup is a $ C_0$-semigroup, see \cite{Yosida.80} p.\ 233.
The generator of a semigroup $ T_t $ is defined as the linear operator $ L $ whose domain $ D(L) $ 
consists of those $ x \in X $ for which there exists a $ y \in X $ such that
\begin{displaymath}
\lim_{t \downarrow 0} {\textstyle \frac{1}{t}}(T_t(x)-x)-y=0
\end{displaymath} 
and the action of $ L $ is then defined by $ L(x)=y $. The limit is taken in the weak topology 
if $ T_t $ is a $ C_0$-semigroup and in the weak-$\ast$-topology in the $ C_0^{\ast}$-case.\\  

We are interested in $ C_0$-semigroups on $ C^{\ast}$-algebras and $ C_0^{\ast}$-semi\-groups
on von Neumann algebras with further properties. First we look at positive semigroups
$ T_t $, i.e.\ semigroups which leave the positive cone $ {\cal A}^+ $ of the algebra invariant:
$ T_t({\cal A}^+)\subseteq {\cal A}^+ $. In Section 3 we assume additionally the unitality $ T_t(\1)=\1 $
and the contractivity of the semigroups. We always assume that the $ C^{\ast}$-algebras possess a unit $ \1 $.\\

In order to handle 
$ C_0$- and $ C_0^{\ast}$-semigroups simultaneously, we use the notation
$ \sigma({\cal A},F) $-continuous semigroup, where 
$ \sigma({\cal A},F) $ is either the locally-convex topology on the $ C^{\ast} $-algebra
$ {\cal A} $ induced by the functionals in the dual $ F={\cal A}^{\ast} $ corresponding to 
$ C_0$-semigroups. 
On the other hand $ {\cal A} $ can be viewed as a von Neumann algebra with predual space 
$ F= {\cal A}_{\ast} $. In this case $ \sigma({\cal A},F) $ is the weak-$\ast$-topology
corresponding
to $ C_0^{\ast}$-semigroups, see \cite{Bratteli.Robinson.87} Section 2.5.3 and Section
3.1.2. For a discussion of the different continuity properties 
on $ C^{\ast} $- and von Neumann algebras, see \cite{Bratteli.Robinson.87} Section 3.2.2.\\

A linear mapping $T: D(T) \to {\cal A} $ on a $C^{\ast}$- and von
Neumann algebra ${\cal A}$ respectively 
with domain $ D(T) \subseteq {\cal A} $
is called symmetric, if $ T(x^{\ast})=T(x)^{\ast} $ for all $ x\in D(T) $.\\

For the theory of $ C^{\ast}$- and
von Neumann algebras the reader is referred to \cite{Bratteli.Robinson.87} and
\cite{Takesaki.79}. References for the theory of one-parameter semigroups are e.g.\ 
\cite{Bratteli.Robinson.87}, \cite{Engel.Nagel.00}, \cite{Pazy.83}.
In \cite{Bratteli.Robinson.87}
the theories of $C_0$- and $C_0^{\ast}$-semigroups are developed together.\\ 

\begin{theorem}\label{theorem1}
Let $ L $ be the generator of a $ \sigma({\cal A},F)$-continuous semi\-group 
$ T_t $ of
symmetric operators on a unital algebra $ {\cal A}$. 
Then the following conditions are equivalent:
\begin{enumerate}
\item $ T_t $ is positive for all $ t \geq 0 $.
\item $ (\lambda -L)^{-1} $ is positive for all large $ \lambda >0 $.
\item $ (\lambda -L)^{-1}(a^2)+a(\lambda -L)^{-1}(\1)a \geq
(\lambda -L)^{-1}(a)a+a(\lambda -L)^{-1}(a) $, for all self-adjoint $ a\in {\cal A} $
and all large $ \lambda > 0$.
\item $ (\lambda -L)^{-1}(\1)+u^{\ast}(\lambda- L)^{-1}(\1)u \geq
(\lambda -L)^{-1}(u^{\ast})u+u^{\ast}(\lambda -L)^{-1}(u) $, for all unitary $u\in {\cal A} $
and all large $ \lambda >0$.
\item $ T_t(a^2)+aT_t(\1)a \geq T_t(a)a+aT_t(a) $, for all self-adjoint $ a\in {\cal A}$ and
$ t\geq 0$.
\item $T_t(\1)+u^{\ast}T_t(\1)u \geq T_t(u^{\ast})u+u^{\ast}T_t(u) $, for all unitary
$u\in {\cal A} $ and $ t\geq 0$.
\end{enumerate}
\end{theorem}

Remark: The proof of the $ C_0^{\ast}$-case requires a modification of the proof of the 
$ C_0$-case in \cite{Bratteli.Robinson.81} . There one uses a convergence argument to prove
the equivalence of $ 5. $ and $ 6. $ to $ 1.-4. $ This argument does not work
in the case of 
$ C_0^{\ast}$-semigroups. In order to overcome this difficulty a 
Laplace transform argument is used in the following proof. 
For the sake of convenience, the complete proof is given.\\

Proof: $ 1.\Rightarrow 2. $: This follows from the Laplace transformation:
\begin{displaymath}
(\lambda-L)^{-1}(a)=\int_0^{\infty}e^{-\lambda t}T_t(a)dt, \qquad \mbox{for all}\quad \lambda > 0,
\quad \mbox{for all}\quad a \in {\cal A}.
\end{displaymath}
The usual relation for $C_0$-semigroups between the resolvent and the Laplace transform holds 
also for
$ C_0^{\ast}$-semigroups, see
\cite{Bratteli.Robinson.87} Proposition 3.1.6.\\
$ 2.\Rightarrow 1. $: 
\begin{displaymath}
T_t(a)= \lim_{n\to\infty}({\textstyle \frac{n}{t}}({\textstyle \frac{n}{t}}-L)^{-1})^n(a),
\qquad \mbox{for all}\quad t>0,
\quad \mbox{for all} \quad a \in {\cal A},
\end{displaymath}
where the limit is taken in the $ \sigma({\cal A},F)$-topology.
Thus, the semigroup is positive whenever the resolvent is positive for $ \lambda $ 
sufficiently large.\\

$1.\Rightarrow3. $ ($ 2.\Rightarrow 4. $):
In the case of uniformly continuous semigroups and bounded generators
respectively conditions $ 1. $ and 
$ 2. $ are equivalent 
to one of the following two conditions \cite{Evans.HancheOlsen.79}:
\begin{eqnarray}
L(a^2)+aL(\1)a &\geq & L(a)a+aL(a), \quad \mbox{for all self-adjoint} \; a\in {\cal A}, \\
L(\1)+u^*L(\1)u &\geq & L(u^*)u+u^*L(u), \quad \mbox{for all unitary} \; u \in {\cal A}.
\end{eqnarray}
With this result in mind, it is sufficient to show the equivalence of 
$ 1. $ and $ 2. $ to the following condition.\\

\em 7.\ The uniformly continuous semigroup $ S_t $ with generator $(\lambda -L)^{-1} $
is positive for all $ t\geq 0$ 
and large $ \lambda >0$.\em\\

$ 2.\Rightarrow 7. $: Follows from
\begin{displaymath} 
S_t=e^{t(\lambda-L)^{-1}}=\sum_{n=0}^{\infty}\frac{t^{n}}{n!}(\lambda-L)^{-n}.
\end{displaymath}

$ 7.\Rightarrow 1. $: Let $ U_t^{\lambda} $ be the uniformly continuous semigroup whose
generator $ L_{\lambda} $
is given by the Yosida ap\-proxi\-ma\-tion of $ L $, i.e. 
$L_{\lambda}:= \lambda L(\lambda-L)^{-1}=\lambda^2(\lambda-L)^{-1}-\lambda $. $ U_t^{\lambda} $
is positive if $ S_t $ is positive, since 
\begin{displaymath}
U_t^{\lambda}=e^{t(\lambda^2(\lambda-L)^{-1}-\lambda)}=e^{-t\lambda}e^{\lambda^2t(\lambda-L)^{-1}}=
e^{-t\lambda}S_{\lambda^2t}.
\end{displaymath}
$ T_t $ is the $ \sigma({\cal A},F)$-limit of $ U_t^{\lambda} $ for $ \lambda \to \infty $,
see \cite{Pazy.83} Theorem 1.5.5.\ for the result in the $ C_0$-case the $ C_0^{\ast}$-result follows
by duality. So $ T_t $ is positive.\\ 

$ 5.\Rightarrow3. $ ($ 6.\Rightarrow 4. $):
We have
\begin{displaymath}
T_t(a^2)+aT_t(\1)a-(T_t(a)a+aT_t(a))\geq 0, 
\end{displaymath}
for all $ t\geq 0 $ and all self-adjoint (unitary $ a\in {\cal A}$). So
\begin{eqnarray*}
& & (\lambda-L)^{-1}(a^2)+a(\lambda-L)^{-1}(\1)a-((\lambda-L)^{-1}(a)a+a(\lambda-L)^{-1}(a))\\
&=& \int_0^{\infty}e^{-\lambda t}(T_t(a^2)+aT_t(\1)a-(T_t(a)a+aT_t(a)) \geq 0
\end{eqnarray*}
follows by Laplace transformation.\\

$ 1. \Rightarrow 5. $ ($ 6. $): Provided $ T_t\geq 0 $ for all $ t\geq 0 $,
\begin{displaymath}
e^{sT_t}=\sum_{n=0}^{\infty}\frac{(sT_t)^n}{n!} 
\end{displaymath}
is positive for all $ s\geq 0$ and $ t\geq 0$. Thus, the estimates (1) and (2) hold 
for the generator $ T_t $ of the semigroup $ e^{sT_t} $.

\section[Positive, unital semigroups of contractions]{Positive, unital semigroups of contractions}

In the previous section we discussed $ \sigma({\cal A},F)$-continuous, positive semigroups.
Now we assume in addition that the semigroup is
unital and contractive. These properties allow
a characterization of the generator which is different to the characterizations
in Theorem \ref{theorem1}. There the characterizations are in terms of the resolvent of the generator,
in the following we provide a characterization in terms of the generator only.

\begin{theorem}\label{theorem2}
Let $ L $ be the generator with domain $ D(L) $ of a $ \sigma({\cal A},F)$-continuous semigroup
$ T_t $ of contractions on $ {\cal A} $.
The following two conditions are equivalent: 
\begin{enumerate}
\item $ T_t $ is positive and unital.
\item $ \1 \in D(L) $, $ L(\1)=0 $ and $ L $ is symmetric.
\end{enumerate}
\end{theorem}

Proof:
$ 1.\Rightarrow 2. $: $ \1 \in D(L) $ and $ L(\1)=0 $ follow from
$ \lim_{h\downarrow 0} \frac{1}{h}(T_h(\1)-\1)=0 $,
where the limit is taken in the $ \sigma({\cal A},F)$-topology.\\

Every element $ x \in {\cal A} $ can be decomposed into a linear combination of four
positive elements $ a, b, c, d \in {\cal A}^+ $ such that
\begin{displaymath}
x=a-b+i(c-d).
\end{displaymath}
Since $ T_t $ is positive and linear we get with the help of this decomposition
\begin{displaymath}
T_t(x^{\ast}) =  T_t(x)^{\ast}, 
\end{displaymath}
thus $ T_t $ is symmetric. Let $ x \in D(L) $, then by using the uniform-continuity of the
$\ast$-map
\begin{eqnarray*}
L(x)^{\ast} & = & \lim_{h \to \infty}{\textstyle \frac{1}{h}}(T_h(x)-x))^{\ast}\\
            & = & \lim_{h \to \infty}{\textstyle \frac{1}{h}}(T_h(x^{\ast})-x^{\ast}),
\end{eqnarray*}
the limit is performed in the $ \sigma({\cal A},F)$-topology.
Thus $ x^{\ast} \in D(L) $ and $ L(x^{\ast})=L(x)^{\ast} $.\\

$ 2. \Rightarrow 1. $: Property $ 2. $ reads in terms of the resolvent 
$ (\lambda -L)^{-1}(\1)=\frac{1}{\lambda}\1$. By approximation in the
$ \sigma({\cal A},F)$-topology we get
\begin{displaymath}
T_t(\1)=\lim_{n\to\infty}({\textstyle \frac{n}{t}}({\textstyle \frac{n}{t}}-L)^{-1})^n(\1)=
\lim_{n\to\infty}({\textstyle \frac{n}{t}})^n({\textstyle \frac{t}{n}})^n(\1)=\1,
\end{displaymath}
which proves the unitality of $ T_t $.\\

By assumption $ L $ is the generator of a semigroup of contractions. Therefore
the range of
$ \lambda - L $ is the whole algebra $ {\cal A} $ for $ \lambda > 0 $. For 
$ x \in {\cal A} $ exists a unique element
$ y \in D(L) $, such that $ (\lambda-L)^{-1}(x)=y $. Clearly, $ \lambda-L $ is symmetric
for $ \lambda > 0 $.
Therefore
\begin{eqnarray*}
((\lambda-L)^{-1}(x))^{\ast} & = & ((\lambda-L)^{-1}(\lambda-L))(y^{\ast})\\
                             & = & (\lambda-L)^{-1}(x^{\ast}).
\end{eqnarray*}
The semigroup $ T_t $ can be defined by 
$ T_t(x)= \lim_{n\to\infty}({\textstyle \frac{n}{t}}({\textstyle \frac{n}{t}}-L)^{-1})^n(x) $.
In the $ C_0^{\ast}$-case the limit is taken in the weak-$\ast$-topology on the von Neumann 
algebra $ {\cal A} $. The weak-$\ast$-topology on $ {\cal A} $ is the relative $ \sigma$-weak
topology on $ {\cal A} $ and the involution $ ^\ast $ is continuous in this topology. So one 
can interchange the involution and the limit, thus $ T_t $ is symmetric.
In the $ C_0$-case the limit is taken in the weak topology on the $C^{\ast}$-algebra 
$ {\cal A} $. Every weakly continuous semigroup is also strongly continuous, see \cite{Yosida.80}
p.233, and the involution $ ^\ast $ is continuous
with respect to the uniform topology on the $ C^{\ast}$-algebra $ {\cal A} $. As in the von
Neumann case we can interchange the limit with the involution and the symmetry of $ T_t $
follows.\\

Now let $  a \in {\cal A}^+ $ and $ a_1:=\frac{a}{\|a\|} $.
$ {\cal A}^+ $ is a convex cone, thus $ a_1 \in {\cal A}^+ $ and 
$ \|a_1\|=1 $. So the spectrum, $ \sigma(a_1) $, of $ a_1 $ is contained in $ [0,1] $.
Therefore 
\begin{displaymath}
1-\sigma(a_1)=\sigma(\1 -a_1)\subseteq [0,1]
\end{displaymath}
and
\begin{displaymath}
\|\1-a_1\| \leq 1.
\end{displaymath}
Thus
\begin{eqnarray*}
\| \1-T_t(a_1)\| & = & \| T_t(\1)-T_t(a_1)\| \\
              & \leq & \|T_t\| \|\1-a_1\| \\
              & \leq & 1 .
\end{eqnarray*}
The semigroup $ T_t $ is symmetric for all $ t\geq 0 $, hence $ T_t(a_1)\in {\cal A}^{sa}$.
With this
\begin{displaymath}
1-\sigma(T_t(a_1))=\sigma(\1-T_t(a_1))\subseteq [-1,1].
\end{displaymath}
It follows that $ \sigma(T_t(a_1))\subseteq [0,2] $, therefore $ T_t(a_1)\in {\cal A}^+ $.
One concludes
\begin{displaymath}
T_t(a)=\|a\|T_t(a_1) \in {\cal A}^+,
\end{displaymath}
thus $ T_t $ is positive, which proves the theorem.\\

Remark: The direction $ 2. \Rightarrow 1. $ can be sharpened using the fact, that a unital,
 contractive map between
two unital $C^{\ast}$-algebras and two von Neumann algebras
respectively is positive, see \cite{Russo.Dye.66}, 
\cite{Bratteli.Robinson.87} Corollary 3.2.6.

\begin{theorem}
Let $ L $ be the generator of a $ \sigma({\cal A},F)$-continuous semigroup $ T_t $ of contractions
on $ {\cal A} $.
\begin{displaymath}
\1 \in D(L), \hspace{0.4em} L(\1)=0 \quad \Rightarrow \quad T_t \hspace{0.4em}\mbox{is positive}.
\end{displaymath}
\end{theorem}

Proof:
The unitality is proven as in Theorem \ref{theorem2}.
Since $ T_t $ is a contraction, i.e. $ \|T_t\|\leq 1 $ for all $ t\geq 0 $, we have $ \|T_t\|=1 $. 
The conclusion now follows from the above remark.

\begin{corollary} 
A unital $ \sigma({\cal A},F) $-continuous semigroup of contractions on an
operator algebra is positive.
\end{corollary}

In the following, the above characterization is restated for the dual and the predual
semigroup respectively.\\
Let $ T_t $ be a $ C_0$-semigroup of contractions on a $ C^*$-algebra $ {\cal A} $.
Then, the adjoint operators $ T_t^* : {\cal A}^*\to {\cal A}^* $ form a $ C_0^*$-semigroup 
of contractions on the dual space $ {\cal A}^* $
of $ {\cal A} $, see \cite{Butzer.Berens.67}.
The generator of $ T_t^* $ is given by the adjoint operator $ L^* $ of
the generator $ L $ of the semigroup $ T_t $. 
In $ {\cal A}^{\ast} $, the set
\begin{displaymath}
{\cal A}^{*+}:= \{ \eta \in {\cal A}^*: \eta(a)\geq 0 \;\; \forall a\in {\cal A}^+\}
\end{displaymath}
is a positive cone \cite{Batty.Robinson.84}.
Let $ \eta\in {\cal A}^{*+} $ and $ T_t $ be a positive $ C_0$-semigroup. Then
$ \eta(T_t(a))\geq 0 $
for all $ a \in {\cal A}^+ $. So $ T_t^* $ is also positive since
$ (T_t^*(\eta))(a)$ $=\eta(T_t(a)) $. For a unital semigroup $ T_t $, one obtains
$ (T_t^*(\eta))(\1)=\eta(\1) $ for all $ \eta \in {\cal A}^* $. $ L(\1)=0 $ implies
$ L^*(\eta)(\1)=0 $ for all $ \eta \in {\cal A} $.\\
In the case that $ T_t $ is a $ C_0^*$-semigroup of contractions on a von
Neumann algebra $ {\cal A} $,
there exists a $ C_0$-semigroup $ T_t^*$ of contractions on the predual $ {\cal A}_* $
which is adjoint to $ T_t $. The unitality and positivity of the semigroup $ T_t $ and
$ L(\1)=0 $ can 
be transferred to $ T_t^* $ and $ L^{\ast} $ as in the case
of a $ C^*$-algebra. Finally, one gets

\begin{corollary}
Let $ T_t^* $ be a $ C_0^*$-semigroup of contractions on $ {\cal A}^* $ the dual of a
$ C^*$-algebra 
$ {\cal A} $ or a $ C_0$-semigroup of contractions on the predual $ {\cal A}_* $ of a von
Neumann algebra $ {\cal A} $. Denote the generator of $T_t^{\ast}$ by $ L^{\ast} $.
The following two conditions are equivalent:
\begin{enumerate}
\item $ T_t^* $ is positive and $ T_t^*(\eta)(\1) =\eta(\1) $ for all $ \eta \in {\cal A}^* $
or for all $ \eta \in {\cal A}_* $.
\item $ L^*(\eta)(\1)=0 $ for all $ \eta \in D(L^*) $.
\end{enumerate}
\end{corollary}

If one applies the last result to ${\cal A}_*=L^1({\cal H}) $ the
Banach space of trace class operators and restricts the semigroup
$ T_t^* $ to the real Banach space $ L^1({\cal H})^{sa} $ of all 
selfadjoint operators in $ L^1({\cal H})$, one sees that the result
due to Kossakowski \cite{Kossakowski.72} is contained in Corollary 2. \\

\noindent{\bf Acknowledgements:} The author would like to thank H.\ Hering, H.\ Leschke and M.\ Sch\"utz. This work
was partially supported by the Deutsche Forschungsgemeinschaft under grant no.\ Le 330/12-1.

\end{document}